\documentclass[apjl]{emulateapj}
\usepackage{apjfonts}
\usepackage{amsmath}
\usepackage{color}

\slugcomment{accepted for publication in ApJL}
\shortauthors{Matsumoto et al.}
\begin{document}
\title{Two-dimensional Numerical Study for Rayleigh-Taylor and Richtmyer-Meshkov Instabilities in Relativistic Jets}
\author{Jin Matsumoto\altaffilmark{1} and Youhei Masada\altaffilmark{2}}
\altaffiltext{1}{Center for Computational Astrophysics, National Astronomical Observatory of Japan, Tokyo, Japan; jin@cfca.nao.ac.jp}
\altaffiltext{2}{Graduate School of System Informatics, Department of Computational Science, Kobe University, Kobe, Japan}
\begin{abstract}
We study the stability of a non-rotating single-component jet using two-dimensional 
special relativistic hydrodynamic simulations. By assuming translational invariance 
along the jet axis, we exclude the destabilization effect by Kelvin-Helmholtz mode. 
The nonlinear evolution of the transverse structure of the jet with a normal jet velocity 
is highlighted. An intriguing finding in our study is that Rayleigh-Taylor and 
Richtmeier-Meshkov type instabilities can destroy cylindrical jet configuration as 
a result of spontaneously induced radial oscillating motion. This is powered by in-situ 
energy conversion between the thermal and bulk kinetic energies. The effective inertia 
ratio of the jet to the surrounding medium $\eta$ determines a threshold for the onset of 
instabilities. The condition $\eta < 1$ should be satisfied for the transverse structure of 
the jet being persisted. 
\end{abstract}
\section{Introduction}
The persistent, well-collimated, relativistic jet is a universal structure that can be observed 
in the astrophysical compact object-accretion disk systems, such as active galactic nuclei 
\citep[AGNs;][]{Begelman84, Ferrari98}, microquasars \citep{Mirabel99}, and potentially,  
gamma-ray bursts \citep[GRBs;][]{Piran04, Meszaros06}. A grand challenge in the 
relativistic physics is to construct a self-consistent theory that is responsible for generation, 
acceleration, and collimation of the astrophysical jet. 

The stability of the relativistic plasma flow is of intrinsic importance in both the acceleration 
and collimation mechanisms of the jet. If the plasma flow propagating relativistically through 
the ambient medium is unstable to internal disturbances, it should be difficult to attain the 
remarkable persistency and velocity of the jet expected in compact object-accretion disk 
systems because of the nonlinear material mixing process caused by instabilities. 

The relativistic jet propagating through the ambient medium is subjected to a storm of instabilities. 
The shear layer, which develops spontaneously at the interface between the jet and the surrounding 
medium, can be destabilized by the Kelvin-Helmholtz instability \citep{Turland76, Blandford76}.
It has been investigated theoretically and numerically in the framework of the astrophysical jet 
\citep{Hardee88, Mizuno07, Perucho10}, and is expected as a primary mechanism of the material 
mixing \citep{Rossi08}. 

In association with the magnetically driven mechanism for the jet-launching \citep{Blandford82, 
Uchida85, Shibata86}, the stability of the magnetized jet is one of the front line topics in  
relativistic physics. The poynting flux-dominated jets carrying large-scale helical magnetic fields 
can become unstable to the kink mode of the current-driven (CD) instability \citep{Lundquist51, 
Spruit 97, Begelman98}. 

A pioneering numerical work done by \citet{Mizuno09} investigated the CD kink instability 
for a static force-free equilibrium by three-dimensional (3D) general relativistic 
magnetohydrodynamic (MHD) simulation (see also Baty \& Keppens 2002 for the case of 
non-relativistic MHD). However, it is still a matter of debate whether the CD kink instability is 
essential for the astrophysical jet because the 3D configuration and strength of the magnetic field, 
which are deeply related to the launching mechanism, are veiled in mystery. 

One of few works that studies the effect of the Rayleigh-Taylor instability \citep[RTI;][]{Rayleigh1900, 
Taylor50} on the jet dynamics is \citet{Meliani07, Meliani09}. They performed 2.5 dimensional numerical 
simulations of two-component jet consisting of a fast inner spine and a slower outer flow. 
They found that the rotational shear that develops at the interfaces between the spine 
and sheath, and the sheath and ambient, plays a key role in triggering the RTI in the 
two-component jet. The driving force of the RTI is, in this case, the centrifugal force. 

Even without the rotation, the relativistic jet potentially becomes unstable to the RTI. 
A radial inertia force naturally arises from a pressure mismatch between the jet and 
surrounding medium when the jet propagates through the ambient medium. The inertia 
force drives the radial oscillating motion of the jet, yielding the reconfinement region 
inside the jet \cite[e.g.,][]{Gomez97, Matsumoto12}. When considering the non-axisymmetric 
evolution of the jet, it might excite the RTI at the interface of the jet (see also \S 3.1). 

In this Letter, the nonlinear development of the relativistic jet is studied using special 
relativistic hydrodynamic (SRHD) simulations. By assuming a non-rotating single 
component jet with translational invariance along the jet axis, the two-dimensional 
stability of the transverse structure of the jet to the radial oscillation-induced RTI is 
investigated. 

\begin{table*}[!htbp]
\begin{center}
\caption{List of Simulation Runs}
\scalebox{1}{\rotatebox{0}{
\begin{tabular}{lcccccccc}
\tableline\tableline
Model & $\rho_{\rm jet,0}c^2$ & $P_{\rm jet,0}$ & $v_{z, \rm jet, 0}/c^2$ & $\gamma_{\rm jet,0}$ & $h_{\rm jet,0}-1$ & $\rho_{\rm ext,0}c^2$ & $P_{\rm ext,0}$ & $\eta_{0}$\\ 
\tableline
A1 & $1 \times 10^{-1}$ & $1 \times 10^{0}$ & 0.99 & 7 & $4 \times 10^{1}$ & $1$ & $1 \times 10^{-1}$ & 147\\ 
A2 & $5 \times 10^{-3}$ & $5 \times 10^{-2}$ & 0.99 & 7 & $4 \times 10^{1}$ & 1 & $5 \times 10^{-3}$ & 10\\
A3 & $5 \times 10^{-4}$ & $5 \times 10^{-3}$ & 0.99 & 7 & $4 \times 10^{1}$ & 1 & $5 \times 10^{-4}$ & 1\\
A4 & $5 \times 10^{-5}$  & $5 \times 10^{-4}$  & 0.99 & 7 & $4 \times 10^{1}$ & 1 & $5 \times 10^{-5}$  & 0.1\\
B1 & $2.5 \times 10^{-1}$ & $2.5 \times 10^{-1}$ & 0.99 & 7 & $4 \times 10^{0}$ & 1 & $2.5 \times 10^{-2}$ & 57\\
B2 & $2.5 \times 10^{-2}$ & $2.5 \times 10^{-2}$ & 0.99 & 7 & $4 \times 10^{0}$ & 1 & $2.5 \times 10^{-3}$ & 6\\
B3 & $2.5 \times 10^{-3}$ & $2.5 \times 10^{-3}$ & 0.99 & 7 & $4 \times 10^{0}$ & 1 & $2.5 \times 10^{-4}$ & 0.6\\
B4 & $2.5 \times 10^{-4}$ & $2.5 \times 10^{-4}$ & 0.99 & 7 & $4 \times 10^{0}$ & 1 & $2.5 \times 10^{-5}$ & 0.06\\
C1 & $1 \times 10^{0}$ & $1 \times 10^{-1}$ & 0.99 & 7 & $4 \times 10^{-1}$ & 1 & $1 \times 10^{-2}$ & 68\\
C2 & $1 \times 10^{-1}$ & $1 \times 10^{-2}$ & 0.99 & 7 & $4 \times 10^{-1}$ & 1 & $1 \times 10^{-3}$ & 7\\
C3 & $1 \times 10^{-2}$ & $1 \times 10^{-3}$ & 0.99 & 7 & $4 \times 10^{-1}$ & 1 & $1 \times 10^{-4}$ & 0.7\\
C4 & $1 \times 10^{-3}$ & $1 \times 10^{-4}$ & 0.99 & 7 & $4 \times 10^{-1}$ & 1 & $1 \times 10^{-5}$ & 0.07\\
D1 & $1 \times 10^{0}$ & $1 \times 10^{-2}$ & 0.99 & 7 & $4 \times 10^{-2}$ & 1 & $1 \times 10^{-3}$ & 52\\
D2 & $1 \times 10^{-1}$ & $1 \times 10^{-3}$ & 0.99 & 7 & $4 \times 10^{-2}$ & 1 & $1 \times 10^{-4}$ & 5\\
D3 & $1 \times 10^{-2}$ & $1 \times 10^{-4}$ & 0.99 & 7 & $4 \times 10^{-2}$ & 1 & $1 \times 10^{-5}$ & 0.5\\
D4 & $1 \times 10^{-3}$ & $1 \times 10^{-5}$ & 0.99 & 7 & $4 \times 10^{-2}$ & 1 & $1 \times 10^{-6}$ & 0.05\\
\tableline\tableline
\end{tabular}}}
\label{table1}
\end{center}
\tablecomments{Columns 9: initial effective inertia ratio of the jet to the surrounding medium defined by Equation (\ref{eq: eta}).}
\end{table*}
\section{Numerical Method} 
\subsection{Basic Equations}
We numerically solve the nonlinear development of a single-component relativistic jet 
in a cylindrical coordinate system $(r, \theta, z)$. By dropping derivatives of the physical 
variables in the z-direction, we exclude the destabilization effect of the Kelvin-Helmholtz 
instability that can grow along the jet direction. We focus simply on the stability of the 
transverse structure of the jet to the non-axisymmetric perturbations. Assuming an ideal 
gas law with a ratio of specific heats $\Gamma = 4/3$, the basic equations are the 
two-dimensional SRHD equations: 
\begin{eqnarray}
\frac{\partial}{\partial t}(\gamma \rho) &+& \frac{1}{r}\frac{\partial }{\partial r}(r \gamma \rho v_{r}) + 
\frac{1}{r} \frac{\partial }{\partial \theta}(\gamma \rho v_{\theta})  = 0 \;, \label{eq: mass conservation} \\ 
\frac{\partial}{\partial t}(\gamma^2 \rho h v_{r}) &+& \frac{1}{r}\frac{\partial }{\partial r} \biggl [ r (\gamma^2 \rho h {v_{r}}^2 + P) \biggr ] \nonumber \\
&+& \frac{1}{r}\frac{\partial }{\partial \theta} (\gamma^2 \rho h v_{r} v_{\theta}) = \frac{P+\gamma^2 \rho h {v_{\theta}}^2}{r} \;, \label{eq: momentum conservation 1} \\
\frac{\partial}{\partial t}(\gamma^2 \rho h v_{\theta}) &+& \frac{1}{r}\frac{\partial }{\partial r} \biggl [ r (\gamma^2 \rho h v_{\theta} v_{r}) \biggr ] \nonumber \\
&+& \frac{1}{r}\frac{\partial }{\partial \theta} (\gamma^2 \rho h {v_{\theta}}^2 + P) = - \frac{\gamma^2 \rho h v_{r} v_{\theta}}{r}\;, \label{eq: momentum conservation 2} \\
\frac{\partial}{\partial t}(\gamma^2 \rho h v_{z}) &+& \frac{1}{r}\frac{\partial }{\partial r} \biggl [ r (\gamma^2 \rho h v_{z} v_{r}) \biggr ] \nonumber \\
&+& \frac{1}{r}\frac{\partial }{\partial \theta} (\gamma^2 \rho h v_{z} v_{\theta}) = 0 \;, \label{eq: momentum conservation 3} \\
\frac{\partial}{\partial t}(\gamma^2 \rho h c^2 - P) &+& \frac{1}{r}\frac{\partial }{\partial r} \biggl [ r (\gamma^2 \rho h v_{r} c^2) \biggr ] \nonumber \\
&+& \frac{1}{r}\frac{\partial }{\partial \theta} (\gamma^2 \rho h v_{\theta} c^2) = 0 \;, \label{eq: energy conservation}
\end{eqnarray}
where $\gamma = 1/\sqrt{1-(v_{r}/c)^2-(v_{\theta}/c)^2 - (v_{z}/c)^2}$ is the Lorentz 
factor and $h = 1 + \Gamma P/ (\Gamma -1)\rho c^2$ is the specific enthalpy. 
The other symbols have their usual meanings. 

A relativistic HLLC scheme is used to solve the SRHD equations 
(\ref{eq: mass conservation})--(\ref{eq: energy conservation}) \citep{Mignone05}. 
The primitive variables are calculated from the conservative variables following the 
method of \citet{Mignone07}. We use a MUSCL-type interpolation method to attain 
second-order accuracy in space while the temporal accuracy obtains second-order by 
using Runge-Kutta time integration. See \citet{Matsumoto12} for more detail on our 
SRHD code. 

The initial rest mass energy density of the external medium and the initial jet velocity 
are common parameters for all the models we studied, and are assumed as 
$\rho_{\rm ext,0}c^2 = 1$ and $(v_r,v_\theta, v_z) = (0,0,0.99c)$, respectively. 
The initial Lorentz factor of the jet is then evaluated as $\gamma_{\rm jet,0} \sim 7$. 
We have three control parameters in our simulations, the initial rest mass energy density 
of the jet ($\rho_{\rm jet,0}c^2$), the initial pressure of the jet ($P_{\rm jet,0}$), and 
the initial pressure of the external medium ($P_{\rm ext,0}$). The fiducial model 
(Model A1 in Table 1) adopts $\rho_{\rm jet,0}c^2 = 0.1$, $P_{\rm jet,0} = 1$ and 
$P_{\rm ext,0} = 0.1$. Note that the jet is initially assumed to be overpressured for all 
the models.

The normalization units in length, velocity, time, and energy density are chosen as 
the initial jet radius $r_{\rm jet,0}$, light speed $c$, light crossing time over the initial 
jet radius $r_{\rm jet,0}/c$, and rest mass energy density in the external medium 
$\rho_{\rm ext,0}c^2$. We use a uniformly spaced grid in cylindrical coordinates 
consisting of $320 \times 200$ zones in $r$- and $\theta$-directions. The computational 
domain spans $0 \le r/r_{jet,0} \le 10$ and $0 \le \theta \le 2\pi$. The initial jet radius is 
resolved by $32$ numeric cells. An outflow (zero gradient) boundary condition is imposed 
on the outer boundary of the domain. The coordinate singularity is treated by placing no 
grid point on the cylindrical axis and filling appropriate "ghost grids" in the region $r < 0$ 
(See, Mohseni \& Colonius 2000 and Ghosh et al. 2010 for the treatment of the coordinate 
singularity with using ghost grids). A separate longer paper will provide our treatment of 
the singular point and its implementation to the HLLC scheme in more detail. By introducing 
small-amplitude ($1\%$) random pressure perturbations to the initial configuration, the 
simulation is initiated.

\section{Results}
\subsection{A Basic Physics Governing Oscillating Motion of Jet}
\begin{figure}[!htbp]
\begin{center}
\scalebox{0.33}{\rotatebox{0}{\includegraphics{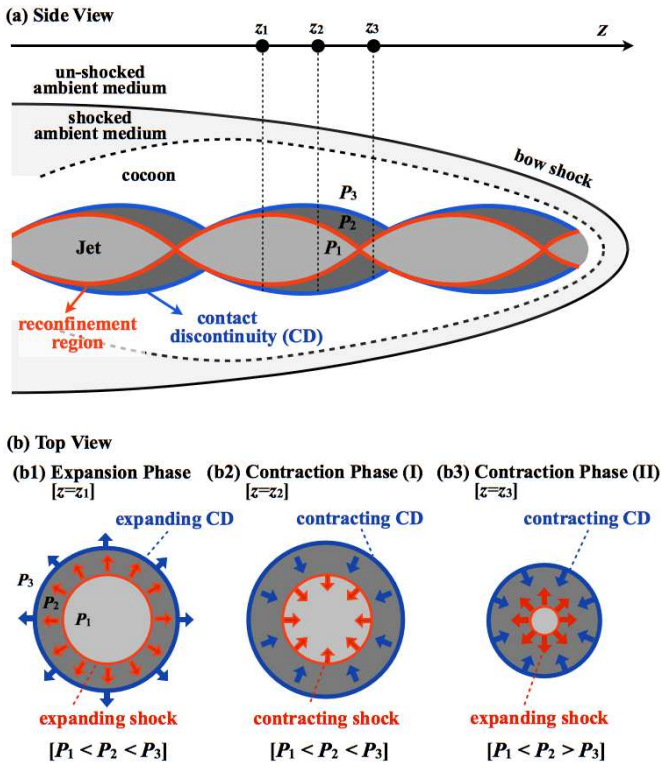}}}
\caption{Schematic picture of the jet propagating through the ambient medium. Top and bottom panels are side and top views of the jet, respectively.}
\label{jet}
\end{center}
\end{figure}
Here, to aid understanding of our numerical results, we briefly describe the basic physics 
governing the radial jet oscillation, which naturally arises from the pressure difference 
between the jet and the external medium. See Matsumoto et al. (2012) for more detail of 
the radial oscillating motion of the jet in the axisymmetric model. 

Figure 1(a) schematically depicts the traditional picture of the relativistic jet propagating 
through the ambient medium. The ambient medium is thermalized at the jet head during 
the jet propagation, and simultaneously forms a cocoon. This is the reason the jet is 
surrounded by the enveloped cocoon material. The transverse structures of the jet 
on cutting planes at $z=z_1$, $z_2$ and $z_3$ are shown in panels (b1)--(b3), 
respectively. The region enclosed by a red curve is the reconfinement region. The blue curve 
denotes the contact discontinuity that separates the jet from the surrounding medium. 

Since the jet is initially overpressured in our model, it starts to expand adiabatically in the 
radial direction (panel (b1)). The adiabatic cooling leads to in-situ energy conversion 
between the thermal and bulk kinetic energies of the jet according to the relativistic 
Bernoulli's principle ($\gamma h \sim {\rm const.}$). The gas pressure inside the jet 
thus becomes smaller than that of surroundings in the expansion phase. The expansion 
is then decelerated by the inward pressure-gradient force acting on the jet-surrounding 
medium interface, and is finally turned into the contraction (panel (b2)). In the subsequent 
phase, the pressure inside the jet increases with the contraction of the jet (panel (b3)). 
The jet restarts to expand radially when the gas pressure of the jet becomes larger than 
that of the surrounding medium.

During the radial oscillation of the jet, the reconfinement region enclosed by a shock surface 
is formed inside the jet \citep{Norman82, Sanders83}. As shown in Figure 1(a), 
the shock-shock collision at the center of the jet excites the outward-propagating shock. 
The collision between the shock and the contact discontinuity results in the contracting 
reconfinement shock. The transition between the outward- and inward-propagating shocks 
occurs repeatedly and shapes the reconfinement region. 

The radial pressure mismatch between the jet and surrounding medium assumed in our 
initial setting is a modeling to reproduce this radial jet oscillation already established in 
axisymmetric models \citep{Daly88, Matsumoto12}. The restoring force of the jet oscillation, 
which is obviously the pressure gradient force acting on the jet-surrounding medium interface, 
might induce the RTI when we consider the non-axisymmetric evolution of the jet. 

\subsection{Time Evolution of the Fiducial Model}
\begin{figure*}[!htbp]
\begin{center}
\scalebox{0.5}{\rotatebox{00}{\includegraphics{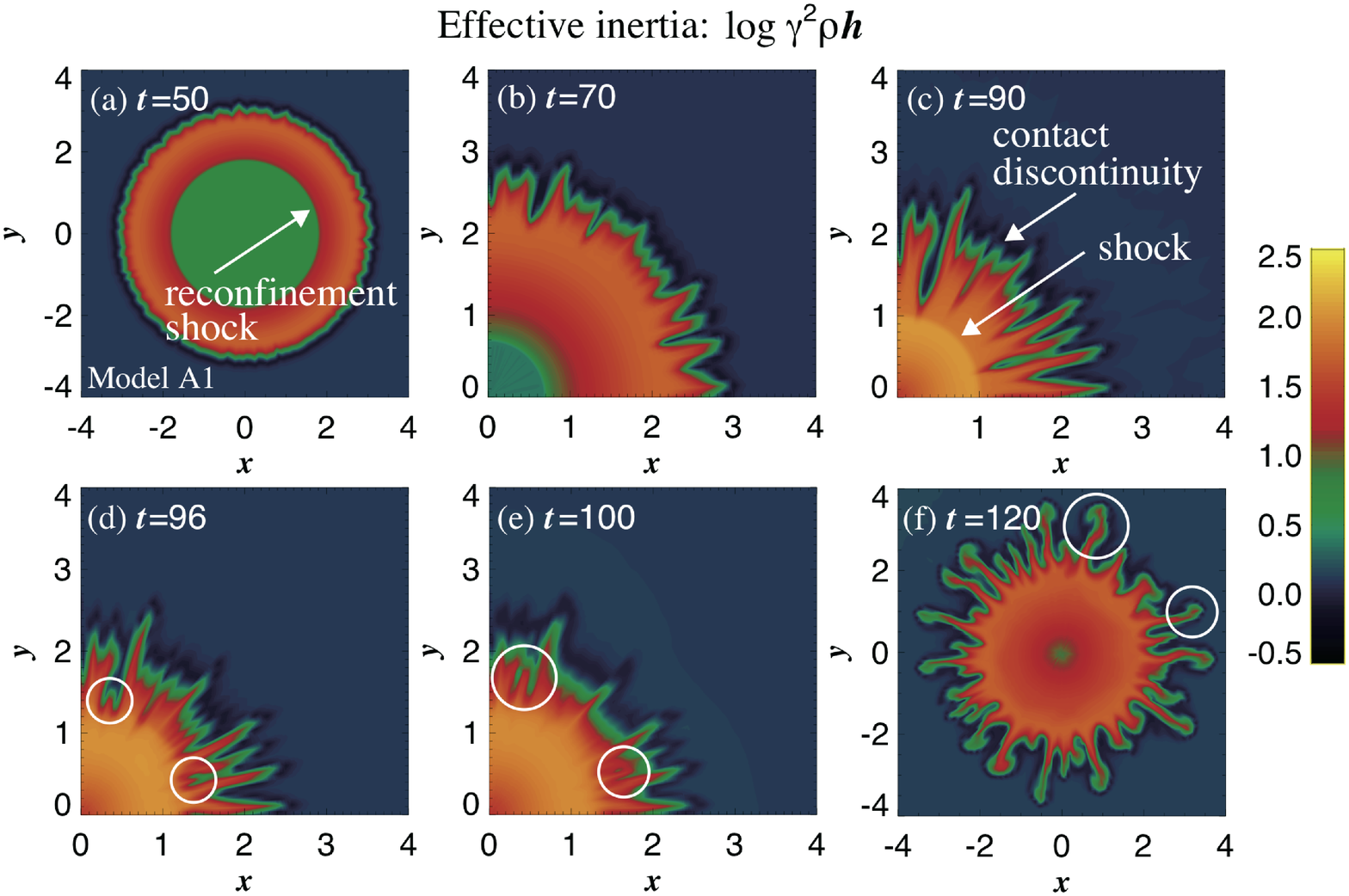}}}
\scalebox{0.85}{\rotatebox{0}{\includegraphics{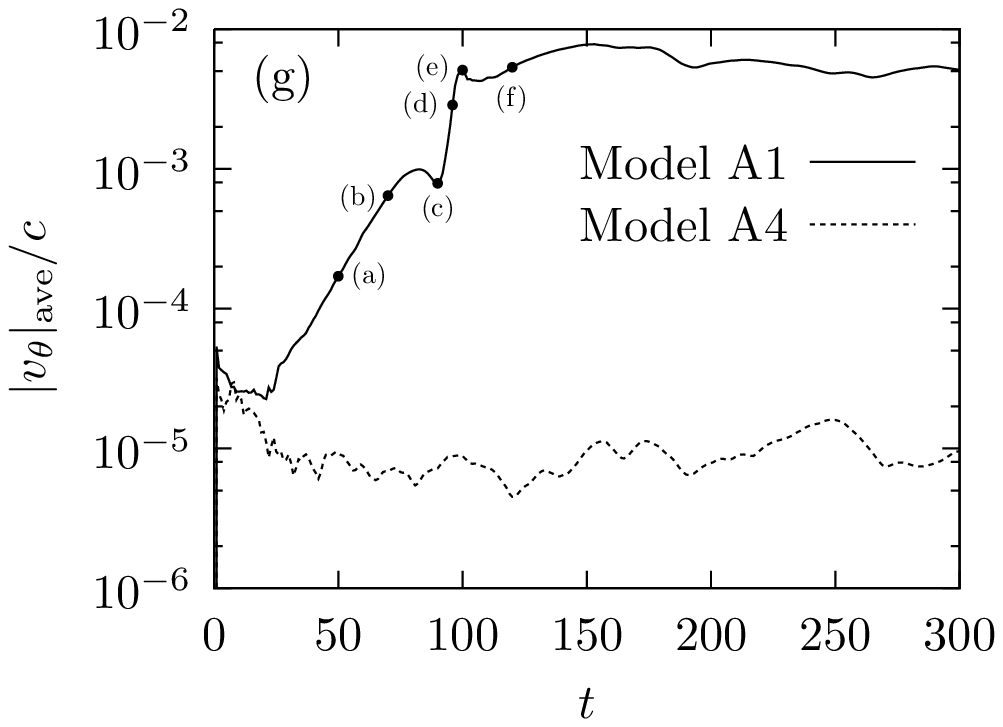}}}
\caption{Panels (a)--(f): Time evolution of the effective inertia $\gamma^2 \rho h$ in the jet-external medium system for the fiducial model A1 in Table 1.
(An animation and a color version of this figure are available in the online journal.) 
Panel (g): Time evolution of the volume-averaged azimuthal velocity $|v_{\theta}|_{\rm ave}$ defined by Equation (\ref{eq: v2ave}).
The filled circles indicate $|v_\theta|_{\rm ave}$ at the time corresponding to the snapshots (a)--(f). 
}
\label{2Dg2roh}
\end{center}
\end{figure*}

Shown sequentially in Figures 2(a)--(f) is the temporal evolution of the cross section 
of the jet for the fiducial model (Model A1 in Table 1). The snapshots (a)--(f) represent 
the transverse distribution of the effective inertia $\gamma^2\rho h$ at $t = 50, 70, 90, 
96, 100$, and $120$, respectively. Figure 2(g) shows the temporal evolution of the 
volume--averaged azimuthal velocity $|v_\theta |_{\rm ave}$ defined by, 
\begin{eqnarray}
|v_{\theta}|_{\rm ave} = \frac{\int_{|v_{z}|>0} \; |v_{\theta}| \; rdr d\theta}{\int_{|v_{z}|>0} \; rdr d\theta} \label{eq: v2ave} \; .
\end{eqnarray}
The solid line denotes the fiducial model. The filled circles indicate $|v_\theta|_{\rm ave}$ 
at the time corresponding to the snapshots (a)--(f). 

The development of the RTI at the interface separating two fluids is controlled by the difference 
in the inertia of the fluid gases in the non-relativistic regime. In the relativistic regime where 
the internal energy of the gas is comparable to or greater than its rest mass energy and/or 
the fluid velocity is relativistic, the inertia of the fluid is enhanced by relativistic effects. 
The effective inertia $\gamma^2\rho h$ is thus a good indicator for studying the gas dynamics, 
especially the stability. Note that the effective inertia of the jet is larger than the external medium 
although the density of the jet is smaller than the external medium in the fiducial model. 

In Figure 2(a), the inward-propagating reconfinement shock is formed behind the corrugated 
contact discontinuity that separates the jet and the external medium. One can find that the 
amplitude of the corrugated jet--external medium interface grows as time passes (Figures 2(a)--(c)). 
Then a finger-like structure, which is a typical outcome of the RTI, emerges in Figure 2(c). 
The RTI is induced by inward pressure gradient force.

The convergence of the inward-propagating reconfinement shock produces 
an outward-spreading shock at the center of the jet. At the timing when the outward going 
shock collides with the contact discontinuity, the Richtmeier-Meshkov instability 
\citep[RMI;][]{Richtmyer60, Meshkov69} is secondarily excited between RTI fingers. 
See Figure 2(d) for the excitation of the RMI fingers that are marked by open circles 
(this process can be clearly seen in the online animation). The evolution of the RMI-driven 
finger marked by the white circle in Figure 2(d) is tracked in Figures 2(e) and (f). We stress 
that almost all finger-like structures in Figure 2(f) have their origin in the RMI.

During the radial oscillating motion of the jet, the two types of finger structures are amplified 
and repeatedly excited at the contact discontinuity, and finally deform the transverse structure 
of the jet. The fiducial model indicates that the transverse structure of the jet is dramatically 
deformed by a synergetic growth of the RTI and RMI once the jet-external medium interface 
is corrugated in the case with the pressure-mismatched jet.

The synergetic growth of the RTI and RMI can be confirmed in Figure 2(g). The volume-averaged 
azimuthal velocity $|v_\theta |_{\rm ave}$ increases exponentially until $t \sim 80$ after 
$t \sim 25$. This is due to the RTI that grows at the jet-external medium interface. 
At around the time $t = 90$ when the outgoing shock passes through the contact 
discontinuity, the evolution property of the $|v_\theta|_{\rm ave}$ is dramatically changed, 
linearly increasing in time. This is evidence of the excitation of the RMI 
because it is well-known that the perturbation amplitude grows linearly with time when 
the RMI develops (Richtmyer 1960; Nishihara et al. 2010). After $t \sim 100$, the system 
enters into the nonlinear saturation stage. 

\subsection{Stability Condition of Jet}
The condition for the transverse structure of the jet being maintained is studied by varying 
two physical parameters that characterize the jet-external medium system. We focus 
on the initial effective inertia ratio between the jet and external medium, which is defined by 
\begin{eqnarray}
\eta_0 = \frac{{\gamma_{\rm jet,0}}^2 \rho_{\rm jet,0} h_{\rm jet,0}}{\rho_{\rm ext,0} h_{\rm ext,0}} \label{eq: eta}
\end{eqnarray}
and the initial specific enthalpy of the jet $h_{\rm jet,0}$ which is a good indicator 
whether the gas is relativistically hot. The model parameters surveyed for examining 
the jet stability are summarized in Table 1 (Models A1--D4). With the same physical and 
computational settings except the parameters $\eta_0$ and $h_{\rm jet, 0}$, we simulate 
various jet-external medium systems. 

In the following, the "unstable model" corresponds to the model in which the transverse 
structure of the jet is deformed in the same way as the fiducial model. In contrast, the "stable model" 
means that the transverse structure of the jet is maintained. As an example of the stable model, 
the temporal evolution of the $|v_\theta |_{\rm ave}$ for Model A4 is demonstrated by 
a dashed line in Figure 2(g). The RTI and RMI are suppressed and the jet interface does 
not deform in the stable model. 

\begin{figure}[!htbp]
\begin{center}
\scalebox{0.85}{\rotatebox{0}{\includegraphics{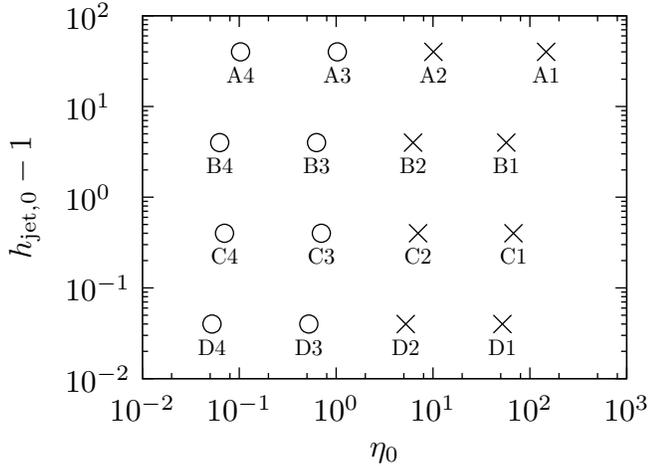}}}
\caption{Stability diagram for the transverse structure of the jet in the $h_{\rm jet,0}$ -- $\eta_0$ parameter space.
The vertical axis represents $h_{\rm jet,0} -1$. Crosses and open circles indicate unstable and stable models listed in Table~1.
}
\label{phase_diagram}
\end{center}
\end{figure}

Figure~\ref{phase_diagram} presents the stability diagram for the transverse structure of the jet in the 
$h_{\rm jet,0}$ -- $\eta_0$ parameter space. The vertical axis shows $h_{\rm jet,0} -1$. The cross and open circle indicate 
the unstable and stable models, respectively. One finds that  the stability criterion 
of the jet can be simply written as $\eta_0 \lesssim 1$ regardless of the specific enthalpy 
of the jet. The jet can maintain its transverse structure as long as the effective inertia 
of the jet is smaller than that of the ambient when excluding the destabilization effects 
by the Kelvin-Helmholtz mode. 

\section{Summary and Discussion} 
In this Letter, the stability of the transverse structure of the non-rotating single-component 
relativistic jet was studied using SRHD simulations. The nonlinear evolution of the 
non-axisymmetric perturbation in the two-dimensional $r$--$\theta$ plane was highlighted 
by assuming a translational invariance along the jet axis. 

Initial pressure mismatch between the jet and surrounding medium results in the  
radial oscillating motion of the jet. An intriguing finding in our study is that the inertia force, 
which acts as a restoring force of the radial oscillating motion,  triggers the primary RTI and 
the secondary RMI at the jet-external medium interface. The transverse structure of the jet is 
remarkably deformed by the nonlinear growth of these radial oscillation-induced instabilities. 

The condition for the transverse structure of the jet being maintained is written, 
regardless of the specific enthalpy of the jet, as $\eta \lesssim 1$, where $\eta$ is the 
effective inertia ratio of the jet to the external medium. This suggests that the inertia ratio 
between the jet and enveloped cocoon material is a critical parameterin a realistic situation, 
as illustrated in figure 1(a), to evaluate the stability of the relativistic jet to the oscillation-induced 
RTI and RMI. 

The condition for the growth of the radial oscillation-induced RTI is the same as that 
for the centrifugally driven RTI found in \citet{Meliani09}. This simply indicates 
that a contact discontinuity separating two fluids with $\eta \gtrsim 1$ becomes unstable 
to the RTI regardless of the origin of the driving force. Unlike the RTI, the growth 
of the RMI at the jet-surrounding medium interface has not been reported in 
\citet{Meliani09} and the other previous studies. The synergetic growth of the RTI 
and RMI, which is a characteristic feature in our unstable model,  
enhances the deformation of the jet interface. 

The growth of the RMI is essentially due to the pressure-mismatched jet that is assumed 
in our initial setting. This induces the radial oscillating motion of the jet and, additionally, 
the RMI at the jet-surrounding medium interface. \citet{Meliani09} adopts the initial condition 
where the radial pressure balance is established in the calculation domain. This would be 
the main reason the RMI does not grow in their work. 

A lot of previous 2D axisymmetric studies suggest that the formation and oscillation 
of the reconfinement shock is a natural result of the jet propagation through the ambient 
medium and is a key to attain the remarkable collimation and persistency of the jet \citep[e.g.,][]
{Marti97, Morsony07, Mizuta09}. However, an important message from our simulations is 
that the non-axisymmetric nature is essential for the stability of the relativistic jet to the 
oscillation-induced instabilities. They would have a great impact on the collimation and 
the acceleration of the three-dimensional jet propagating through ambient medium. The 
high-resolution realistic 3D simulation of the jet propagation is within the scope of 
our work and will be reported in our subsequent paper. 
\acknowledgments
We thank Kazunari Shibata, Hiroyuki R. Takahashi, and Akira Mizuta for thier useful discussions.
Numerical computations were carried out on Cray XT4 and the general-purpose PC farm at the
Center for Computational Astrophysics, the National Astronomical Observatory of Japan 
and on SR16000 at  YITP at Kyoto University.
\appendix
\section{Ability of HLLC and HLL Schemes to Simulate Instabilities}
The evolution of the RTI and RMI was simulated in this Letter. To capture these 
instabilities, a numerical scheme capable of accurately resolving a contact discontinuity is 
required. In our simulations, we use a relativistic HLLC scheme \citep{Mignone05}, 
which is a modification of the HLL scheme \citep{Harten83} by restoring the missing 
contact wave in the solution of the Riemann problem. Here we shed light on the ability 
of the HLL and HLLC schemes to simulate these instabilities using codes developed by 
\citet{Matsumoto11} and \citet{Matsumoto12}, respectively.

\begin{figure}[!htbp]
\begin{center}
\scalebox{0.5}{\rotatebox{0}{\includegraphics{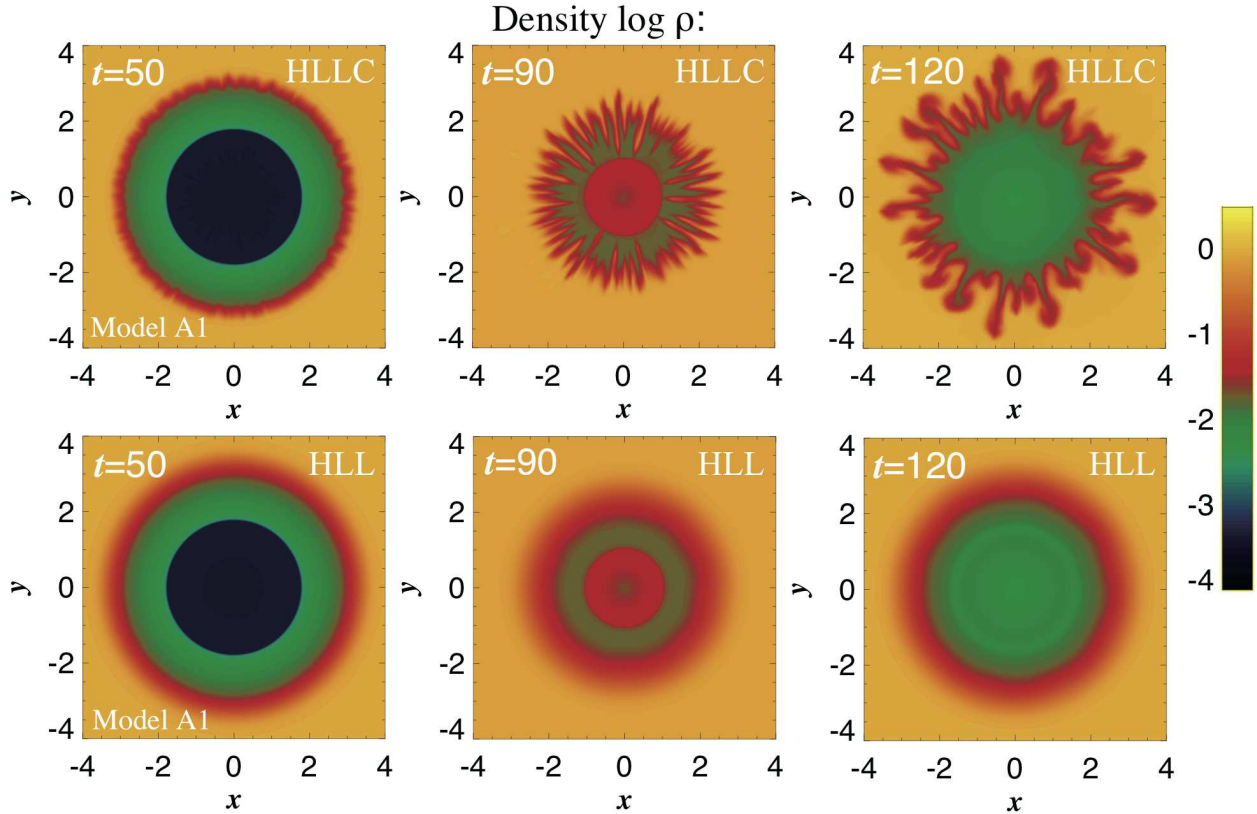}}}
\caption{Density contours of the cross section of the jet for the fiducial model A1 in Table~1 when $t=50$, $90$, and $120$.
Upper and lower panels correspond to the case evolved by HLLC and HLL schemes, respectively.
(A color version of this figure is available in the online journal.)}
\label{hllc_vs_hll}
\end{center}
\end{figure}

Figure~\ref{hllc_vs_hll} shows the temporal evolution of the cross section of the jet for 
the fiducial model (Model A1 in Table 1). The upper and lower panels place, in order of 
time, results obtained in models evolved by HLLC and HLL schemes, respectively. It is 
surprising that the RTI and RMI fingers do not emerge in the model with HLL scheme 
although the completely same initial settings and grid spacings ($320\times 200$ zones 
in $r-$~and $\theta-$directions) are adopted in both models. With our simulation codes, 
three times higher resolution is needed to capture the RTI and RMI in the HLL scheme 
than in the HLLC scheme. 


\end{document}